\documentclass[aps, prb, twocolumn, superscriptaddress,floatfix]{revtex4}

\usepackage{bm}
\usepackage{graphicx}
\usepackage{amsmath}
\usepackage{color}
\usepackage{soul}

\newcommand{\bk}{\bm{k}}

\newcommand{\diff}{\mathrm{d}}
\newcommand{\eps}{\varepsilon}

\newcommand{\im}{\mathrm{i}}

\newcommand{\kdotp}{ \bm{k} \cdot \bm{p} }
\newcommand{\kdottp}{ {\bm{k} \cdot \tilde{\bm{p}}} }

\newcommand{\OsloC}{Centre for Materials Science and Nanotechnology,  Department of Physics, University of Oslo,  Norway}

\newcommand{\bp}{\bm{p}}
\newcommand{\br}{\bm{r}}
\newcommand{\cmm}{\,\text{cm}^{-3}}

\newcommand{\dFD}{\frac{\partial f_{\rm FD}(\varepsilon)}{\partial \varepsilon}}

\newcommand{\VASP}{\textsc{VASP}}

\begin{document}

\title{Thermoelectric transport of  GaAs, InP, and PbTe:  Hybrid functional with $\kdottp$ interpolation versus scissor-corrected generalized gradient approximation}

\author{Kristian Berland}   
\email{kristian.berland@smn.uio.no}
 \affiliation{\OsloC}
\author{Clas Persson} \affiliation{\OsloC} 

\begin{abstract}
Boltzmann transport calculations
based on band structures generated with density functional theory (DFT) are 
often used in the discovery and analysis of thermoelectric materials. 
In standard implementations, such calculations require dense $\bk$-point sampling of the Brillouin zone
and are therefore typically limited to the generalized gradient approximation (GGA),
whereas more accurate methods such as hybrid functionals would have been preferable.
GGA variants, however, generally underestimate the band gap.
While premature onset of minority carriers 
can be avoided with scissor corrections, the band gap also affects the band curvature.
In this study, we resolved the $\bk$-point sampling issue in hybrid-functional based calculations by extending our recently developed $\kdottp$ interpolation scheme [Comput. Mater. Sci. 134, 17 (2017)] 
to non-local one-electron potentials and spin-orbit coupling. 
The Seebeck coefficient generated based on hybrid functionals were found to
agree better than GGA with experimental data for GaAs, InP, and PbTe.
For PbTe, even the choice of hybrid functional has bearing on the interpretation of experimental data,
which we attribute to the description of valley convergence of the valence band. 
\end{abstract}
\maketitle

\section{Introduction}

By converting heat to electricity, 
thermoelectrics can recover some of the immense waste heat
generated in transport, power generation, and industrial processes.
The aim of reducing global CO$_2$ emissions and
recent record thermoelectric conversion efficacies\cite{thermoe:review} have both led to renewed interest in 
optimizing and uncovering new thermoelectric materials. 
Such efforts are aided by the Boltzmann transport equation (BTE) based on first-principle density functional theory (DFT) calculations.
For instance, the doping concentration that maximizes the product of 
electrical conductivity $\sigma$ and the Seebeck coefficient squared  $S^2$, i.e. thermoelectric power factor $\sigma S^2$, 
can be easily estimated by shifting the Fermi level $\mu_{\rm F}$ in the calculations. 
A large power factor and low  thermal conductivity $\kappa$ is essential for obtaining a large figure of merit $ZT = \sigma S^2 T/ \kappa $. 
With DFT-generated band structure, non-parabolicity and multi-valley are included from the onset,
which is key in
computational screening of thermoelectric materials.\cite{Carrete:Nanograined,Xi2016:rational,zhang2016:designing,Chen2016:TrendsWithExp,Madsen:automated,ricci_ab_2017} 
However, such calculations are limited by the need for dense sampling of the Brillouin zone.\cite{boltztrap} 
This is not a major issue when employing semi-local exchange-correlation (XC) functionals in the generalized gradient approximation (GGA),\cite{PBE} but more sophisticated approaches such as hybrid functionals\cite{PBE0,HSE06} and GW\cite{GW:review,GWmethod} are typically out of reach.

Many thermoelectric materials including PbTe have narrow direct band gaps,
which leads to large non-parabolicity and small effective masses 
-- in a two-band Kane model, the effective mass is given by $m = (3\hbar^2/4P^2) E_{\rm gap}$, 
where $P$ is the momentum matrix coupling between the two bands.\cite{Sofo1994:optimum} 
Thermoelectric power factors
can therefore be optimized by effectively tuning the band gap using solid solutions, strain, or native vacancies.\cite{Band_engineering} 
While scissor corrections can be used to avoid premature onsets of minority carrier transport  
caused by GGA's underestimated band gaps,\cite{Berland2016:filtering,example_of_scissor} this approach does not correct effective masses and non-parabolicity.\cite{Persson:2001,PERSSON2006}  
Unlike for dielectric functions,\cite{LinOpt,Del_sole:1993} there is no formal justification for using scissor corrections for transport properties.
While underestimated gaps can be avoided 
with certain GGAs,\cite{AK13,EngelVosko93} meta-GGAs,\cite{TranBlaha2009} and related schemes\cite{GLLB1,GLLB2} designed to improve band gaps at low computational costs, more computationally demanding methods, such as hybrid functionals and GW,
can be desirable for describing broad classes of solids. 
The same hybrid functional can also be employed for optimizing the structure as for computing the band structure. 
Moreover, perturbative methods, generally do not correct inaccurate electronic densities which could arise for materials incorrectly found to be semi-metals 
and may not resolve spurious band hybridizations. 
For transport calculations, the more computationally demanding methods can be employed if the band structure is accurately interpolated.
In particular, interpolation methods that make use of information in the Kohn-Sham (KS) wave functions, 
such as the Shirley,\cite{Shirley,Predegast:Shirley} or the Wannier\cite{wannier2001,wannier90new,WannierReview,BoltzWann} method can be efficient.
Recently, we developed the $\kdottp$ method,\cite{berland201717} 
which corrects the standard extrapolative $\kdotp$ method\cite{Persson2007280}
by ensuring  that the $\kdotp$ and KS energies match on a coarse mesh.
In this work, we extended the scheme to hybrid functionals and spin-orbit coupling by replacing momentum-matrix elements by velocity-matrix elements.

The next section illustrates the need for a dense $\bk$-mesh in BTE calculations using \textsc{BoltzTraP}\cite{boltztrap} 
and the effectiveness of our interpolation method.
In Sec. III, the thermoelectric transport properties of hybrid functionals and scissor-corrected  GGA are compared for the direct band-gap III-V semiconductors GaAs, InP and the cubic chalcogenide PbTe.
While III-V semiconductors are not traditional thermoelectric materials, they are important in electronics.
Here they serve as model systems as GGA significantly underestimates their band gap energies 
and accurate experimental Seebeck coefficient are available for a range of doping concentrations. 
PbTe, on the other hand, is a narrow-gap thermoelectric material of significant scientific and commercial interest.
In the final section we offer our perspective on prospects of enhancing the accuracy and predictability of first-principles calculations 
of thermoelectric properties. 
\section{Method}

\subsection{Density functional theory calculations}
The atomic and electronic structure were obtained using 
DFT with the projector-augmented plane wave code \VASP.\cite{vasp1,vasp3,vasp4,vasp:optics}
Three different functionals were considered for obtaining the electronic structure:
the GGA functional PBE,\cite{PBE} the screened hybrid HSE06\cite{HSE06} (hereafter HSE) and the hybrid PBE0.\cite{PBE0} These calculations relied on a $12\times12\times12$ $\bm{k}$-point sampling of the $\bk$-point grid and 64 bands in total with spin-orbit coupling included.   

For each of these three functionals, the atomic structure was kept the same to solely explore 
the effect of the XC functional on the band structure and exclude effects arising from differing lattice parameters this.
A proper comparison of contributions from lattice parameter relaxation would require taking thermal expansion into account for each of the functionals and this is beyond the scope of this paper. 
The atomic structures\footnote{Initial structures  are obtained from the \textsc{Materials Project}.\cite{MatProject}} employed were 
relaxed using the GGA functional PBEsol\cite{PBEsol} with a $10\times10\times10$ $\bm{k}$-point sampling and an energy cutoff set 30\%\ above the standard high energy cutoff.

\subsection{Corrected $\kdotp$-based interpolation scheme}
\label{sec:kdottp}

Developed in the 1950s, the $\kdotp$ method\cite{KANE1956,LuttingerKohn1955,dresselhaus2007group} has become a
standard method to represent electronic band structures. 
It has found wide usage in the envelope-function formalism,\cite{Jeongnim1998,Allmen1992,Stanko:kp}
which extends the theory to semiconductor heterostructures.
The $\kdotp$ Hamiltonian at a given $\bk$ is given in terms of  Bloch wave function basis of a specific $\bk_0$, i.e. $\psi_{i,\bk_0}(\br) = u_{i,\bk_0}(\br) e^{\im \bk_0\cdot \br}$.
The Hamiltonian is given by  
\begin{align}
H_{ij}(\bk) =  \left( \varepsilon_{i,\bk_0} +  \frac{\hbar^2 \left(\bk - \bk_0 \right)^2}{2 m}\right) \delta_{ij}  +  \frac{\hbar (\bk -\bk_0)\cdot \bm{p}_{ij} }{m}\,,\label{eq:H}
\end{align}  
where $\varepsilon_{i,\bk_0}$ are the electron eigenvalues at $\bk_0$ and $\bm{p}_{ij} = \langle \psi_{i,\bk_0} | \hat{\bp} | \psi_{j,\bk_0} \rangle$ are the corresponding momentum matrix elements. 
Traditionally, the $\kdotp$ Hamiltonian is parameterized, for a finite number of bands,
based on a combination of measured and calculated properties,\cite{Dresselhaus1955:kp,Cardona1966,voon2009k} 
leading to, for instance, two-, eight- and thirty- band models etc.\cite{voon2009k} 
Here we avoid this empiricism by computing the matrix elements directly using DFT.

The $\kdotp$ Hamiltonian is an exact representation of the Schr\"odiger equation for local potentials, $V(\br)$, in the limit of infinite number of bands.
\begin{figure}[t!]
\includegraphics[width=8cm]{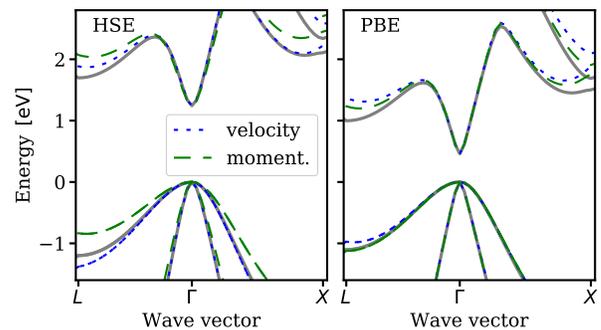}
\caption{Calculated band structure of GaAs. The panels compare the KS result generated with HSE (left panel) and PBE (right panel) 
with the 15-band $\kdotp$ result, with momentum-  (green, long-dashed) and velocity-  (blue short-dashes) matrix elements calculated at the $\Gamma$-point.
\label{fig:kp_variants}}
\end{figure}
However, practical calculations are based on a finite number of bands and plane-wave DFT calculations often rely on non-local pseudopotentials, both contribute to making the $\kdotp$ Hamiltonian inexact.
In our previous work,\cite{berland201717}  we employed the harder  so-called ``GW'' projector-augmented wave (PAW) pseudopotentials to enhance the accuracy of the $\bm{p}_{ij}$ matrix elements.\cite{vasp1,vasp3,vasp4} 
Truly local potentials can be obtained with all-electron codes or exact PAW calculations. 
However, for hybrid functionals, the Fock exchange part is intrinsically non-local.
To extend the $\kdotp$ method, and hence our interpolation scheme, to non-local potentials, we replaced the momentum matrix elements $\bm{p}_{ij}$ by velocity (i.e. conjugate momentum)  matrix elements  $ m \bm{v}_{ij}  =  \frac{\im}{\hbar} \langle \psi_{i,\bk_0} | [H, \bm{r}] | \psi_{j,\bk_0} \rangle$. This choice makes the $\kdotp$ Hamiltonian (\ref{eq:H}) exact to second order in $(\bk-\bk_0)$.\cite{Pickard2000}
Moreover, these matrix elements include spin-orbit coupling if this was employed at the DFT level.
The velocity matrix elements were obtained by adapting  \VASP\cite{vasp1,vasp3,vasp4} routines for calculating dielectric function in the longitudinal gauge.\cite{vasp:optics}  
With velocity, rather than momentum matrix elements, we found no appreciable advantage in using the harder ``GW pseudopotentials'', and we therefore employed standard PAW pseudopotentials in this work.

Figure~\ref{fig:kp_variants} compares the HSE (left panel) and PBE (right panel) band structure of GaAs with the $\kdotp$ band structure generated with 15 bands using respectively momentum (long dashes) and velocity matrix elements (short dashes). Spin-orbit coupling is not included in the comparison as we were only able to extract it with velocity-matrix elements.
The figure shows that for HSE, velocity matrix elements provide far better agreement with the KS band structure than momentum matrix elements. 
For PBE, both agree well with the KS band structure, with momentum matrix elements providing a slightly better agreement, 
which could in part be due to error cancellations, such as the use of a finite number of bands.

Depending on the solid, we found that the $\Gamma$-point based $\kdotp$ method does not generally agree as well with with the KS band structure as in Fig.~\ref{fig:kp_variants}.
With increasing number of bands,
the $\kdotp$ method should approach the exact KS band structure for local potentials and for non-local potentials, the leading error due
due to band truncation $\propto(\bk-\bk_0)^2$ should vanish. 
However, in practice the $\kdotp$ band structures can converge slowly with number of bands and at some point, stop improving due to numerical inaccuracies in the matrix elements.
To obtain an accurate interpolated band structure in the entire Brillouin zone, extrapolating from multiple $\bk_n$ points can therefore be an attractive option. 
Moreover, in typical DFT calculations of solids, the KS wave functions are available for a finite number of $\bk$ points, as needed to converge the electronic density.
In  extrapolating from multiple $\bk$ points, however, discontinuities arise at the meeting points of two extrapolations. Extrapolating from $\bk_0$ and 
$\bk_1$, we would in general find that $\eps^{\kdotp}_{\bk_0} (\bk) \neq \eps^{\kdotp}_{\bk_1} (\bk)$, where the subscript denotes the wave vector used to construct the $\kdotp$ Hamiltonian.
One option is simply to average between different $\kdotp$ results; to exemplify, $\overline{\eps^{\kdotp}}(\bk) = a \eps^{\kdotp}_{\bk_0}(\bk) + (1-a) \varepsilon^{\kdotp}_{\bk_1}(\bk)$. However, as the errors arising in two extrapolations often share sign, wiggles, and spurious band extrema can arise.

In the $\kdottp$ method, a correction term is introduced that ensures that the $\kdotp$ energies 
generated by extrapolating from, for instance, $\bk_0$ matches the neighbouring
$\bk$ points in the coarse KS mesh,  i.e. $\eps^\kdottp_{\bk_0}(\bk_1) = \eps^{\rm KS}(\bk_1)$. 
This correction term is akin to Kane parameters in few-band $\kdotp$ models, accounting for bands not included explicitly in the $\kdotp$ Hamiltonian. 
In a three-dimensional reciprocal space, for the case of $\bk$ values enclosed by a tetrahedron with corners at $\bk_0$ (reference point), $\bk_1$, $\bk_2$, and $\bk_3$ (target points), 
the correction term takes the form:
 \begin{align}
\Delta H(\bk) =   \frac{m}{\hbar} \sum_{n=1,2,3}  \Omega_n(\bk)   \frac{(\bk - \bk_0)^2}{(\bk_n - \bk_0)^2 }\sum_i \delta \varepsilon_{i,\bk_n} 
V_{i,\bk_n} V_{i,\bk_n}^\dagger\,. \label{eq:Hcorr}
\end{align}
Here $V_{i,\bk_n}$ are the eigenvectors of Eq.\,(\ref{eq:H}) for $\bk$ values at the respective target points $n=1,2,3$. $V_{i,\bk_n} V_{i,\bk_n}^\dagger$ are band projections which account for band crossings and changing band nature, whereas
$\Omega_n(\bk)$ is an angular projection term,  given by  
\begin{align}
  \Omega_n(\bk)  &= \frac{\left[ \bm{s}_n \cdot  \left( \bk - \bk_0 \right) \right]^2 }{ \sum_{n}  \left[ \bm{s}_n \cdot  \left( \bk - \bk_0 \right) \right]^2}\,,
\end{align}
with 
$ \bm{s}_1 = \left(\delta \bk_2 \times\delta \bk_3 \right)/\left[ \delta \bk_1 \cdot ( \delta \bk_2 \times \delta \bk_3 ) \right]$ where $\delta \bk_i = \bk_i -\bk_0$ and similarly for $\bm{s}_2$  and $\bm{s}_3$.
For sake of symmetry, since several of $\kdottp$ extrapolations starting from the coarse mesh ends up at same dense mesh points, the method averages over 
different $\kdottp$ results, which results slight smoothening of the results.\cite{berland201717}

\subsection{Boltzmann transport equation} 

The Boltzmann transport equation (BTE) in the relaxation-time approximation
is commonly used to study the electronic transport in solids. 
In this approach, the conductivity $\sigma$ and Seebeck coefficient $S$, can be expressed in terms of 
the transport-spectral function $\Sigma(\eps)$, as follows (tensor indices suppressed),
\begin{align}
\sigma &=  e^2 \int_{-\infty}^\infty \diff \eps\, \left(-  \dFD \right) \Sigma(\eps)\,,  \\
T\sigma S &= e  \int_{-\infty}^\infty \diff \eps\,\left( - \dFD \right)\Sigma(\eps) (\eps-\mu_{\rm F})\,. \label{eq:Seebeck}
\end{align}  
Here, the Fermi window $ \left( - \dFD \right)$  is given by the derivative of the Fermi-Dirac distribution function.
The transport spectral function is given by 
\begin{align} 
\Sigma(\epsilon) &=\frac{1}{V N} \sum_{\bk, i}\,v_i(\bk) v_i(\bk) \tau_i(\bk) \, \delta\left( \eps-\varepsilon_i(\bk) \right)\,,
\end{align}
where $v_i(\bk)$ are the group velocities.
In our study, $\Sigma(\eps)$ was computed with \textsc{BoltzTraP}\cite{boltztrap} based on  $\kdottp$  interpolated  band structures. 
As the purpose of our study was to probe the effect of the band structure account, we simply employed the constant-relaxation time approximation $\tau_i(\bk) = \tau$.  

\begin{figure} [t!]
  \includegraphics[width=8cm]{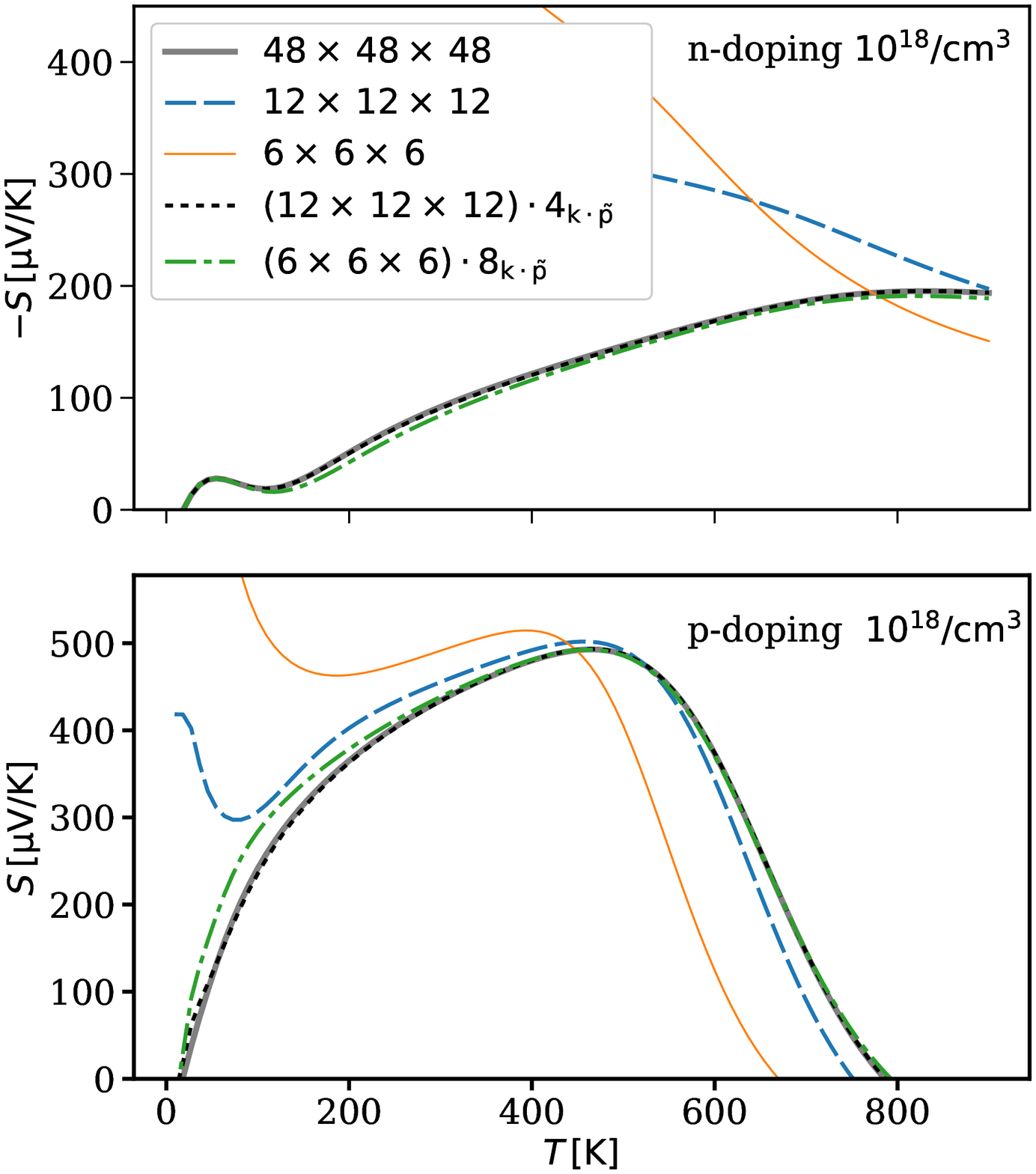}
  \caption{Seebeck coefficient of GaAs as function of temperature calculated for different $\bk$-meshes, comparing standard and $\kdottp$ based results.  The upper panel shows result for n-doping and the lower, p-doping of $10^{18}\cmm$. 
Here $(12\times12\times12)_{4_\kdottp}$ denotes that the $12\times12\times12$ mesh was interpolated to a $48\times48\times48$ mesh.  
\label{fig:ConvSeebeck}
}
\includegraphics[width=8cm]{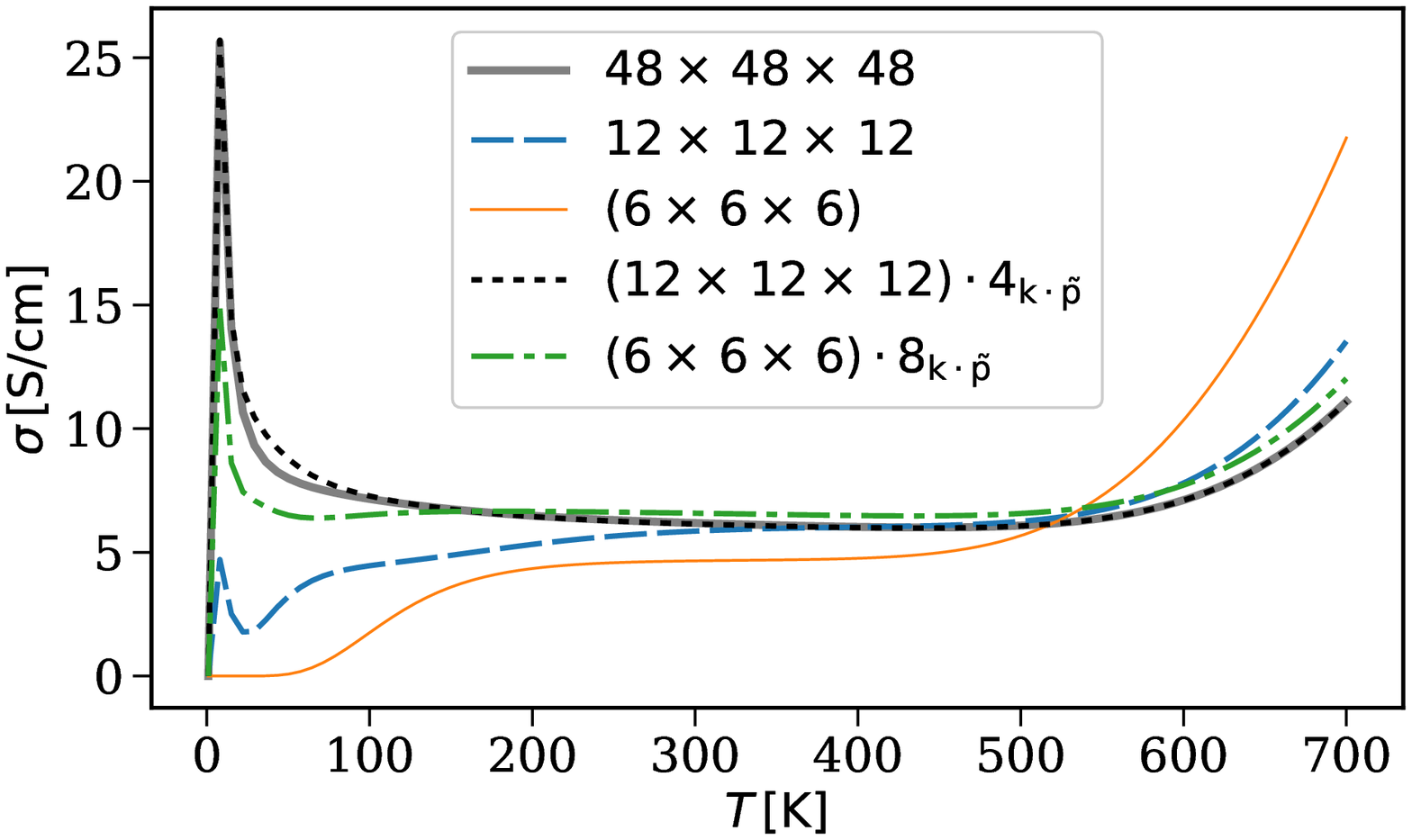}
\caption{Conductivity of GaAs with p-doping as in Fig.~\ref{fig:ConvSeebeck}. \label{fig:pdos_gaas}  }
\end{figure}

\begin{figure} 
\includegraphics[width=8cm]{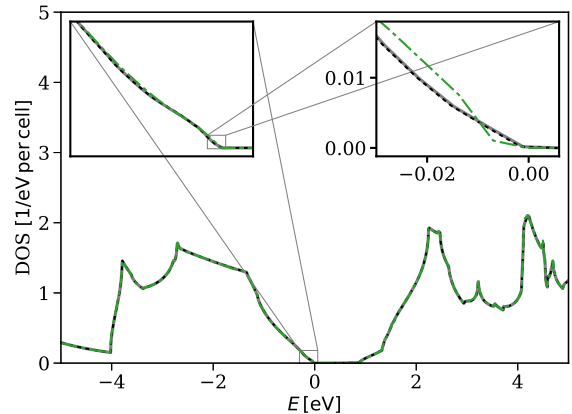}
\caption{Density of states of GaAs generated with the tetrahedron method using a  $48\times48\times48$ mesh based on standard KS results (full grey curve) and $\kdottp$ interpolation from a  $12\times12\times12$ mesh (dashed black)  and  a $6\times 6\times 6$ mesh (dash-dotted green).   The insets each scale the figure by a factor of 10. \label{fig:dos_gaas} }
\end{figure}

\subsection{Convergence study}

Our  $\bk$-mesh convergence study for the Seebeck coefficient of GaAs is shown in Fig.~\ref{fig:ConvSeebeck}
with n- and p- carrier densities set to $10^{18}\cmm$  (upper and lower panel, respectively),
whereas Fig.~\ref{fig:pdos_gaas} shows the corresponding convergence study for the conductivity in the p-doped case. 
The thin orange, blue dashed, and thick gray  curves are generated with respectively $6\times6\times6$, $12\times12\times12$, and 
$48\times48\times48$ $\bk$-meshes. 
The green dashed-dotted and black curves with short dashes are based on the KS solutions of a $6\times6\times6$ and $12\times12\times12$ mesh, both  interpolated using the $\kdottp$ method to a $48\times48\times48$ mesh.
At 200 K, the interpolated Seebeck coefficient based on a  $12\times12\times12$  mesh deviates 
from the reference $48\times48\times48$ KS result by no more than 1.4\% for n  and 0.65\% for p doping. 
The deviation is generally slightly larger at lower temperatures and smaller at higher temperatures. 
Interpolating from a $6\times6\times6$ mesh, the deviations increase to 17\%  (3.6\%)  for n (p) doping, which is too inaccurate in a quantitative analysis, but deviations are modest at high temperatures. 
Fig.~\ref{fig:ConvSeebeck} shows that the interpolation from a $12\times12\times12$, but not from  $6\times6\times6$ mesh, is able to resolve the spike in the conductivity at 8 K,  with a peak value differing by no more than 0.08/$\mathrm{\Omega} \text{cm}$ from the reference KS results. 
This result can be linked to how well the interpolation is able to able to resolve the density of states very close to the conduction band edge, as shown in Fig.~\ref{fig:dos_gaas}. 
At this low temperature, the transport is completely dominated by holes close to the band edge. 
The figure shows that while interpolating from a $6\times6\times6$ is sufficient to resolve the overall DOS, interpolating from $12\times12\times12$ mesh is needed to recover the DOS with high precision close to the band edge.  

While accuracy can be enhanced by increasing the number of empty bands, we found that interpolating a $12\times12\times12$ $\bk$-mesh using a convenient number of bands, at least twice that of the number of occupied bands, generally gives a highly accurate interpolation of the near-gap conduction and valence band states contributing to the thermoelectric transport. 
The $\kdottp$ method can be extended to very dense grids at low computational cost. The results in the next section are based on an interpolation to a $60\times60\times60$ mesh for GaAs and InP and to a $96\times96\times96$ mesh for PbTe. 

\section{Results  \label{sec:results}}

\subsection{GaAs and InP: n-doping}

While bulk GaAs and InP are poor thermoelectric materials, nanowires of these compounds could serve as highly efficient thermoelectric devices.\cite{InP:nanowire,GaAs:nanowire}
The transport properties of GaAs and InP are also of interest in high-frequency electronics, light emission and detection, and photovoltaics. 
Owing to their high material quality and dopability, they are well suited as test systems for exploring how using different XC functional affects thermoelectric transport properties. 

\begin{figure}[h]
\includegraphics[width=8cm]{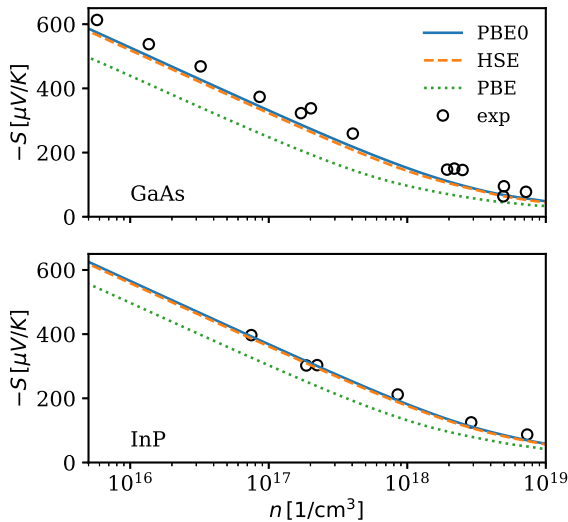}
\caption{Pisarenko plot of the Seebeck coefficient of n-doped GaAs and InP  at 300~K comparing theory and experiment.  
Experimental data for GaAs as given by Pichanusakorn\cite{GaAs:Seebeck} and references therein, and for InP by Kesamanly~et~al.\cite{InP:Kesamanly} as reported by Rode.\cite{GaAs:Seebeck}
}
\label{fig:GaAs_doping}
\end{figure}
A Pisarenko plot comparing the experimental\cite{GaAs:Seebeck,InP:Kesamanly} and calculated Seebeck coefficients of n-doped GaAs and InP at 300~K is shown in Fig.~\ref{fig:GaAs_doping}.
For all doping concentration, the two hybrids HSE and PBE0 give similar results, which agree better with experiments than PBE. 
The better agreement for the two hybrids can be linked to more accurate band gaps resulting in a more accurate band curvature close to the conduction band minimum, as shown in 
table~\ref{tab:gaps}. 
The calculated effective masses were here estimated with the standard $\kdotp$ method with a small, rather than infinitesimal, $|\delta \bk|$ of $0.01/\text{\AA}$, as spin orbit coupling distorts the band structure close to the $\Gamma$ point.  

\begin{table}[h]
\caption{Band gap and effective mass of GaAs and InP. \label{tab:gaps}}
\begin{ruledtabular}
\begin{tabular}{lcc}
&    GaAs  &       InP \\
Method  &    $\Delta_\Gamma [{\rm eV}] $ ($m_c$) &  $\Delta_\Gamma [{\rm eV}]$ ($m_c$)  \\
\hline
PBE &  0.38 (0.028)   &  0.50 (0.043)  \\
HSE &  1.13 (0.063)   &  1.23  (0.074)  \\
PBE0 &  1.64  (0.063) &  1.75  (0.081)   \\
Exp.\footnotemark[1] &  1.519 (0.067)  &  1.4236  (0.0795) \\
\end{tabular}
\end{ruledtabular}
\footnotetext[1]{Experimental band gap energies and effective masses are based on the recommended values of Vurgaftman~et~al.\cite{Vurgaftman2001}.}
\end{table}

In the non-degenerate limit of a n-doped single parabolic band model at constant scattering time, the Seebeck coefficient can be expressed as follows,\cite{nolas2013thermoelectrics} 
\begin{align}
    S &=  - \frac{k_B}{e} \left[  \frac{5}{2}  - \log\left( \frac{n/2}{ \left( 2 \pi  m_e k_{\mathrm{B}}T \right)^{3/2}} \right)  \right]\,.
    \label{eq:Sebeck_non}
\end{align}
The logarithmic dependence on the carrier density explains why the three methods give similar, virtually linear, slopes up to about $10^{18} \cmm$ in Fig.~\ref{fig:GaAs_doping}.
In this formula, using two different masses $m_1$ and $m_2$ results in constant shift in the Seebeck coefficient, as follows
\begin{align}
  S_{\rm m_1} -    S_{\rm m_2} = \frac{3}{2}\log(\frac{m_2}{m_1})\,. 
  \label{eq:Seebeck_shift}
\end{align}
This shift agrees fairly well with the difference in offset in Fig.~\ref{fig:ConvSeebeck}.
To exemplify, at a doping concentration of  $5 \cdot 10^{15}\cmm$, the Seebeck coefficient of PBE is shifted $100~\mu {\rm V/K}$ compared to HSE, while Eq.~(\ref{eq:Seebeck_shift}) gives  105 $\mu{\rm V/K}$  with effective masses provided in table~\ref{tab:gaps}. 
At higher doping concentrations, the Seebeck coefficients become more similar, which is related both to non-parabolicity and the fact the electrons become degenerate. 

\begin{figure}[t!]
\includegraphics[width=8cm]{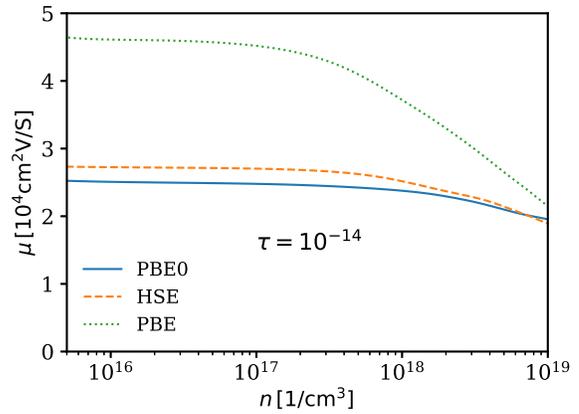}
\caption{Mobility of GaAs calculated with a constant relaxation time $\tau = 10^{-14} {\rm s}$. 
\label{fig:GaAs_mobility}}
\end{figure}

Figure~\ref{fig:GaAs_mobility} shows that the choice of XC functional also has a strong effect on the mobility $\mu  = \sigma/e n$ at a constant relaxation time. That the mobility is 1.7 larger for PBE than HSE at a doping concentration of $5 \cdot 10^{15}\cmm$ is similar to, but somewhat smaller than $m_{\rm PBE}/m_{HSE} = 2.3$.
The larger change in mobility with doping concentration for PBE than HSE and PBE0 can be linked to the increased  non-parabolicity at smaller bandgaps. 
In some studies, the relaxation time is fitted to the measured mobility at a given temperature and doping concentration.
This fitting can veil the role of the XC functional, but this would not mimic the role of non-parabolicity at higher temperatures or doping concentrations. 
As we employed a constant relaxation time, none of the methods are even close to describing measured drops in mobility (not shown in figure), which is about a factor of five from   $5 \cdot 10^{15}\cmm$ to $10^{19}\cmm$.\cite{Sotoodeh:2000} 
The large drop can be linked to the important role of impurity scattering arising from dopants, which in turn could also explain the underestimated Seebeck coefficient even when employing hybrid functionals\cite{GaAs:Seebeck}

\subsection{PbTe}

PbTe has been used in radioisotope thermoelectric generators, powering for instance the Viking space probes.\cite{PbTe:review}
Even if it was one of the first thermoelectric materials discovered, PbTe and related chalcogenides have continued to attract scientific interest, with new record ZTs being reported the last decade.\cite{Heremans554,Lee:rigid_band,jaworski:alloy,Pei:heavy_hole,pei2011convergence}
In 2008 Heremans~et~al.\cite{Heremans554} measured a ZT of 1.5 at 773 K, which was achieved through 2\% Tl substitution of the Pb site.
The high efficiency has been attributed to the formation of resonant states in the valence band.\cite{Heremans554,Lee:rigid_band,jaworski:alloy,Pei:heavy_hole}
Using the Engel-Vosko (EV) GGA functional,\cite{EngelVosko} which avoids the severe band gap underestimates of standard GGAs, 
Singh\cite{Singh:PbTe} calculated the Seebeck coefficient of PbTe. 
However,  even if doping was only accounted for 
implicitly by shifting the Fermi level, beyond a p-carrier concentration of $10^{19}\cmm$, 
the results agreed better with the high Seebeck coefficient measured by Heremans.~et~al 
than the lower Seebeck coefficient reported by Airapetyants~et~al.,\cite{airapetyants1966structure} and others for several different dopants.\cite{Pei:heavy_hole} This finding motivated us to re-examine  the Seebeck coefficient using hybrid functionals.

\begin{figure}[t!]
\includegraphics[width=8cm]{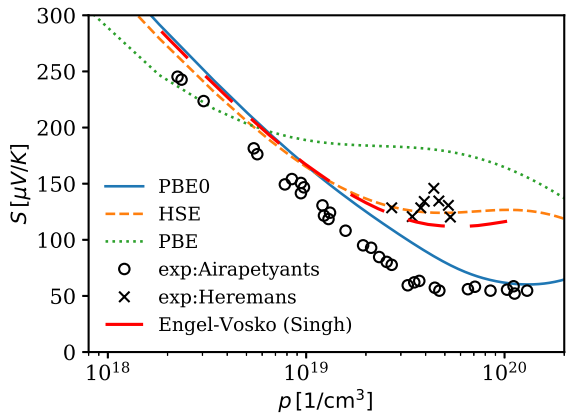}
\caption{The Seebeck coefficient of PbTe as a function of p-carrier concentration at 300 K.
Circles represents experimental data of Airapetyants~et~al.,\cite{airapetyants1966structure} 
while crosses correspond to Heremans et al.\cite{Heremans554}
}
\label{fig:PbTe_p}
\end{figure}
Figure~\ref{fig:PbTe_p} compares the Seebeck calculated based on PBE, HSE, and PBE0 with the experimental measurements of Airapetyants~e.~al.\cite{airapetyants1966structure} and Heremans~et~al.,\cite{Heremans554} as presented by Pei.~et~al.\cite{Pei:heavy_hole}
For carrier concentrations less than about $5\cdot 10^{18}\cmm$, both hybrids give similar results and slightly overestimate the experimental Seebeck cofficient. 
While the magnitude agrees with experiment, the slope of the Seebeck coefficient with PBE becomes inaccurate beyond $2\cdot 10^{18}\cmm$.
At low doping concetrations, the difference between PBE and the hybrids can be linked to the band curvature of the $L$ point valence-band maximum, 
which in turn is related to the band gap. The PBE band gap is 0.065 eV, wheres that of HSE is 0.42 eV and that of PBE0 is 0.96 eV.
As the experimental band gaps is 0.2 eV at 0~K and 0.3~eV at room temperature,\cite{PbTe:temp_gap} 
PBE severely underestimates the gap, but PBE0 considerably overestimates it.
At a doping concentration of about $5\cdot 10^{19}\cmm$, PBE overestimates the Seebeck coefficient by close to a factor of four, 
whereas HSE agrees well with EV calculations of Singh,\cite{Singh:PbTe} and the measurements of Heremans~et~al.\cite{Heremans554} 
PBE0, on the other hand, agrees well with the results of Airapetyants~et~al.,\cite{airapetyants1966structure} 
and is therefore in line with earlier investigations attributing the high Seebeck coefficient of Heremans~et~al. to the inclusion of Tl impurities.\cite{Heremans554,Lee:rigid_band,jaworski:alloy,Pei:heavy_hole}

\begin{figure}[h!]
\includegraphics[width=8cm]{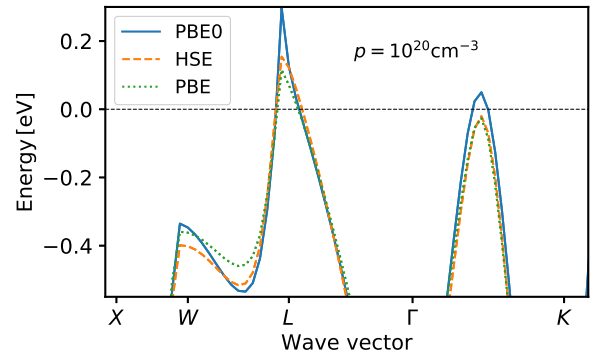}
\caption{Band structure of PbTe. The Fermi level is set to a p-carrier concentration of $2\cdot10^{19}\cmm$ 
\label{fig:PbTe_p_bas}}
\end{figure}

The valence band structure of PbTe calculated with the three different methods, as shown in Fig.~\ref{fig:PbTe_p_bas},
casts some light on the difference between HSE and PBE0 at higher doping concentration.
The Fermi level was set to yield a p-carrier concentration of $10^{20}\cmm$ for each of the three XC functionals, as the Seebeck coefficients of the XC functionals differ strongly in this region. 
A noticeable feature of the PbTe band structure is the approximate valley convergence of the $L$ maximum and the maximum along the $\Sigma$ line between $\Gamma$ and $K$. This valley convergence is one of the reasons why PbTe exhibits a high power factor.\cite{pei2011convergence} 
While the $\Sigma$ valley of HSE is non-degenerate at this doping concentration, it becomes degenerate with PBE0. 
The $L$ valley of of PBE0 is also narrower than that of HSE. Both these features contribute to reducing the Seebeck coefficient of PBE0 compared to HSE, 
but they also enhance the conductivity, which we found to be 1.4 times larger for PBE0 than HSE at this doping concentration. 
 
\begin{figure}[h!]
\includegraphics[width=8cm]{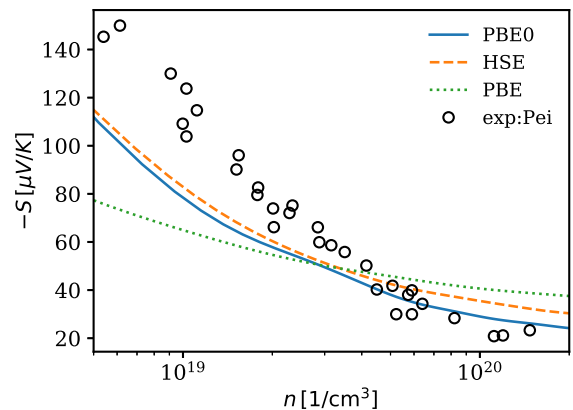}
\caption{Seebeck coefficient of PbTe as a function of n-carrier concentration at 300 K. The circles represent the experimental data of
  Pei et~al.\cite{Pei:PbTE_ndoping}
  \label{fig:PbTe_n}}
\end{figure}

Finally, Fig~\ref{fig:PbTe_n} compares the calculated Seebeck coefficient of n-doped PbTe with experimental data by Pei.~et~al.\cite{Pei:PbTE_ndoping}  
Similar to the case of GaAs, both hybrids give similar results and significantly improves the agreement with experiment, but the calculated Seebeck coefficient is lower than the measured. 

\section{Outlook}

First-principles methods, such as DFT,  regularly provide highly accurate predictions for a range of material properties, 
including structural, energetic, and optical properties.
However, for electronic transport in thermoelectric materials, DFT-based studies are generally limited to qualitative analysis or material screening.  
In this paper, we have shown that the direct band gap underestimation of standard GGAs is one source of inaccuracy in such calcualtions, as the band gap is strongly linked to the band curvature of conduction and valence bands. 
However, even if we can expect hybrid functionals to improve the accuracy, in particular for the Seebeck coefficient, 
other effects must also be accounted for to realize truly predictive thermoelectric transport calculations. 
Scattering mechanisms should also be described based on  first-principles calculations and the effect of thermal expansion and explicit doping on the band structure should be included.

Even in screening studies, when a quantitative comparison between theory and experiments is not sought, our study illustrates 
pitfalls of using standard GGA. Some materials might be deemed more favorable than others due to inaccurate effective masses caused by band gap underestimation. 
Moreover, valley convergence, one of the most promising ways of optimizing electronic transport properties, can be furtitious and arise due to computational choises. 
We have shown that even if the hybrids HSE and PBE0 provide more accurate band structures than GGAs, 
the hybrids provide somewhat differing accounts of the valley convergence, which in turn affects the Seebeck coefficient. 
We also note that while hybrid functionals generally improve the description of electron quasiparticle band gaps compared to GGAs, 
their ability to accurately describe the full band structure in the near gap region, and thus transport spectral function $\Sigma(\epsilon)$, has not been widely assessed.

Finally, we have shown that by improving the recent $\kdottp$ method, we can accurately interpolate the band structure and thus overcome the $\bk$-point sampling issue 
of hybrid functional based transport calculations. 
Our study should motivate using wave-function based interpolation methods in combination with more advanced first-principles methods in future screening studies, or as input in machine learning, of thermoelectric material properties.

 \section{acknowledgement}

We thank Ole Martin L\o vvik and Espen Flage-Larssen for discussions. 
Computations were performed on the Abel and Stallo high performance cluster through a NOTUR allocation. This work is part of THELMA project (Project No. 228854) supported by the Research Council of Norway.

\bibliography{library} 

\begin{thebibliography}{68}
\expandafter\ifx\csname natexlab\endcsname\relax\def\natexlab#1{#1}\fi
\expandafter\ifx\csname bibnamefont\endcsname\relax
  \def\bibnamefont#1{#1}\fi
\expandafter\ifx\csname bibfnamefont\endcsname\relax
  \def\bibfnamefont#1{#1}\fi
\expandafter\ifx\csname citenamefont\endcsname\relax
  \def\citenamefont#1{#1}\fi
\expandafter\ifx\csname url\endcsname\relax
  \def\url#1{\texttt{#1}}\fi
\expandafter\ifx\csname urlprefix\endcsname\relax\def\urlprefix{URL }\fi
\providecommand{\bibinfo}[2]{#2}
\providecommand{\eprint}[2][]{\url{#2}}

\bibitem[{\citenamefont{Zhang and Zhao}(2015)}]{thermoe:review}
\bibinfo{author}{\bibfnamefont{X.}~\bibnamefont{Zhang}} \bibnamefont{and}
  \bibinfo{author}{\bibfnamefont{L.-D.} \bibnamefont{Zhao}},
  \bibinfo{journal}{J. Materiomics} \textbf{\bibinfo{volume}{1}},
  \bibinfo{pages}{92 } (\bibinfo{year}{2015}).

\bibitem[{\citenamefont{Carrete et~al.}(2014)\citenamefont{Carrete, Mingo,
  Wang, and Curtarolo}}]{Carrete:Nanograined}
\bibinfo{author}{\bibfnamefont{J.}~\bibnamefont{Carrete}},
  \bibinfo{author}{\bibfnamefont{N.}~\bibnamefont{Mingo}},
  \bibinfo{author}{\bibfnamefont{S.}~\bibnamefont{Wang}}, \bibnamefont{and}
  \bibinfo{author}{\bibfnamefont{S.}~\bibnamefont{Curtarolo}},
  \bibinfo{journal}{Adv. Funct. Mater.} \textbf{\bibinfo{volume}{24}},
  \bibinfo{pages}{7427} (\bibinfo{year}{2014}).

\bibitem[{\citenamefont{Xi et~al.}(2016)\citenamefont{Xi, Yang, Wu, Yang, and
  Zhang}}]{Xi2016:rational}
\bibinfo{author}{\bibfnamefont{L.}~\bibnamefont{Xi}},
  \bibinfo{author}{\bibfnamefont{J.}~\bibnamefont{Yang}},
  \bibinfo{author}{\bibfnamefont{L.}~\bibnamefont{Wu}},
  \bibinfo{author}{\bibfnamefont{J.}~\bibnamefont{Yang}}, \bibnamefont{and}
  \bibinfo{author}{\bibfnamefont{W.}~\bibnamefont{Zhang}}, \bibinfo{journal}{J.
  Materiomics} \textbf{\bibinfo{volume}{2}}, \bibinfo{pages}{114 }
  (\bibinfo{year}{2016}).

\bibitem[{\citenamefont{Zhang et~al.}(2016)\citenamefont{Zhang, Song, Madsen,
  Fischer, Zhang, Shi, and Iversen}}]{zhang2016:designing}
\bibinfo{author}{\bibfnamefont{J.}~\bibnamefont{Zhang}},
  \bibinfo{author}{\bibfnamefont{L.}~\bibnamefont{Song}},
  \bibinfo{author}{\bibfnamefont{G.~K.} \bibnamefont{Madsen}},
  \bibinfo{author}{\bibfnamefont{K.~F.} \bibnamefont{Fischer}},
  \bibinfo{author}{\bibfnamefont{W.}~\bibnamefont{Zhang}},
  \bibinfo{author}{\bibfnamefont{X.}~\bibnamefont{Shi}}, \bibnamefont{and}
  \bibinfo{author}{\bibfnamefont{B.~B.} \bibnamefont{Iversen}},
  \bibinfo{journal}{Nat. Commun.} \textbf{\bibinfo{volume}{7}}
  (\bibinfo{year}{2016}).

\bibitem[{\citenamefont{Chen et~al.}(2016)\citenamefont{Chen, Pohls, Hautier,
  Broberg, Bajaj, Aydemir, Gibbs, Zhu, Asta, Snyder
  et~al.}}]{Chen2016:TrendsWithExp}
\bibinfo{author}{\bibfnamefont{W.}~\bibnamefont{Chen}},
  \bibinfo{author}{\bibfnamefont{J.-H.} \bibnamefont{Pohls}},
  \bibinfo{author}{\bibfnamefont{G.}~\bibnamefont{Hautier}},
  \bibinfo{author}{\bibfnamefont{D.}~\bibnamefont{Broberg}},
  \bibinfo{author}{\bibfnamefont{S.}~\bibnamefont{Bajaj}},
  \bibinfo{author}{\bibfnamefont{U.}~\bibnamefont{Aydemir}},
  \bibinfo{author}{\bibfnamefont{Z.~M.} \bibnamefont{Gibbs}},
  \bibinfo{author}{\bibfnamefont{H.}~\bibnamefont{Zhu}},
  \bibinfo{author}{\bibfnamefont{M.}~\bibnamefont{Asta}},
  \bibinfo{author}{\bibfnamefont{G.~J.} \bibnamefont{Snyder}},
  \bibnamefont{et~al.}, \bibinfo{journal}{J. Mater. Chem. C}
  \textbf{\bibinfo{volume}{4}}, \bibinfo{pages}{4414} (\bibinfo{year}{2016}).

\bibitem[{\citenamefont{Madsen}(2006)}]{Madsen:automated}
\bibinfo{author}{\bibfnamefont{G.~K.~H.} \bibnamefont{Madsen}},
  \bibinfo{journal}{J. Am. Chem. Soc.} \textbf{\bibinfo{volume}{128}},
  \bibinfo{pages}{12140} (\bibinfo{year}{2006}).

\bibitem[{\citenamefont{Ricci et~al.}(2017)\citenamefont{Ricci, Chen, Aydemir,
  Snyder, Rignanese, Jain, and Hautier}}]{ricci_ab_2017}
\bibinfo{author}{\bibfnamefont{F.}~\bibnamefont{Ricci}},
  \bibinfo{author}{\bibfnamefont{W.}~\bibnamefont{Chen}},
  \bibinfo{author}{\bibfnamefont{U.}~\bibnamefont{Aydemir}},
  \bibinfo{author}{\bibfnamefont{G.~J.} \bibnamefont{Snyder}},
  \bibinfo{author}{\bibfnamefont{G.-M.} \bibnamefont{Rignanese}},
  \bibinfo{author}{\bibfnamefont{A.}~\bibnamefont{Jain}}, \bibnamefont{and}
  \bibinfo{author}{\bibfnamefont{G.}~\bibnamefont{Hautier}},
  \bibinfo{journal}{Sci. Data} \textbf{\bibinfo{volume}{4}},
  \bibinfo{pages}{170085} (\bibinfo{year}{2017}).

\bibitem[{\citenamefont{{Madsen} and {Singh}}(2006)}]{boltztrap}
\bibinfo{author}{\bibfnamefont{G.~K.~H.} \bibnamefont{{Madsen}}}
  \bibnamefont{and} \bibinfo{author}{\bibfnamefont{D.~J.}
  \bibnamefont{{Singh}}}, \bibinfo{journal}{Comput. Phys. Commun.}
  \textbf{\bibinfo{volume}{175}}, \bibinfo{pages}{67} (\bibinfo{year}{2006}).

\bibitem[{\citenamefont{Perdew et~al.}(1996{\natexlab{a}})\citenamefont{Perdew,
  Burke, and Ernzerhof}}]{PBE}
\bibinfo{author}{\bibfnamefont{J.~P.} \bibnamefont{Perdew}},
  \bibinfo{author}{\bibfnamefont{K.}~\bibnamefont{Burke}}, \bibnamefont{and}
  \bibinfo{author}{\bibfnamefont{M.}~\bibnamefont{Ernzerhof}},
  \bibinfo{journal}{Phys. Rev. Lett.} \textbf{\bibinfo{volume}{77}},
  \bibinfo{pages}{3865} (\bibinfo{year}{1996}{\natexlab{a}}).

\bibitem[{\citenamefont{Perdew et~al.}(1996{\natexlab{b}})\citenamefont{Perdew,
  Ernzerhof, and Burke}}]{PBE0}
\bibinfo{author}{\bibfnamefont{J.~P.} \bibnamefont{Perdew}},
  \bibinfo{author}{\bibfnamefont{M.}~\bibnamefont{Ernzerhof}},
  \bibnamefont{and} \bibinfo{author}{\bibfnamefont{K.}~\bibnamefont{Burke}},
  \bibinfo{journal}{J. Chem. Phys.} \textbf{\bibinfo{volume}{105}},
  \bibinfo{pages}{9982} (\bibinfo{year}{1996}{\natexlab{b}}).

\bibitem[{\citenamefont{Heyd et~al.}(2003)\citenamefont{Heyd, Scuseria, and
  Ernzerhof}}]{HSE06}
\bibinfo{author}{\bibfnamefont{J.}~\bibnamefont{Heyd}},
  \bibinfo{author}{\bibfnamefont{G.~E.} \bibnamefont{Scuseria}},
  \bibnamefont{and}
  \bibinfo{author}{\bibfnamefont{M.}~\bibnamefont{Ernzerhof}},
  \bibinfo{journal}{J. Chem. Phys.} \textbf{\bibinfo{volume}{118}},
  \bibinfo{pages}{8207} (\bibinfo{year}{2003}).

\bibitem[{\citenamefont{Onida et~al.}(2002)\citenamefont{Onida, Reining, and
  Rubio}}]{GW:review}
\bibinfo{author}{\bibfnamefont{G.}~\bibnamefont{Onida}},
  \bibinfo{author}{\bibfnamefont{L.}~\bibnamefont{Reining}}, \bibnamefont{and}
  \bibinfo{author}{\bibfnamefont{A.}~\bibnamefont{Rubio}},
  \bibinfo{journal}{Rev. Mod. Phys.} \textbf{\bibinfo{volume}{74}},
  \bibinfo{pages}{601} (\bibinfo{year}{2002}).

\bibitem[{\citenamefont{Aryasetiawan and Gunnarsson}(1998)}]{GWmethod}
\bibinfo{author}{\bibfnamefont{F.}~\bibnamefont{Aryasetiawan}}
  \bibnamefont{and}
  \bibinfo{author}{\bibfnamefont{O.}~\bibnamefont{Gunnarsson}},
  \bibinfo{journal}{Rep. Prog. Phys.} \textbf{\bibinfo{volume}{61}},
  \bibinfo{pages}{237} (\bibinfo{year}{1998}).

\bibitem[{\citenamefont{Sofo and Mahan}(1994)}]{Sofo1994:optimum}
\bibinfo{author}{\bibfnamefont{J.~O.} \bibnamefont{Sofo}} \bibnamefont{and}
  \bibinfo{author}{\bibfnamefont{G.~D.} \bibnamefont{Mahan}},
  \bibinfo{journal}{Phys. Rev. B} \textbf{\bibinfo{volume}{49}},
  \bibinfo{pages}{4565} (\bibinfo{year}{1994}).

\bibitem[{\citenamefont{Pei et~al.}(2012{\natexlab{a}})\citenamefont{Pei, Wang,
  and Snyder}}]{Band_engineering}
\bibinfo{author}{\bibfnamefont{Y.}~\bibnamefont{Pei}},
  \bibinfo{author}{\bibfnamefont{H.}~\bibnamefont{Wang}}, \bibnamefont{and}
  \bibinfo{author}{\bibfnamefont{G.~J.} \bibnamefont{Snyder}},
  \bibinfo{journal}{Adv. Mater.} \textbf{\bibinfo{volume}{24}},
  \bibinfo{pages}{6125} (\bibinfo{year}{2012}{\natexlab{a}}).

\bibitem[{\citenamefont{Berland et~al.}(2016)\citenamefont{Berland, Song,
  Carvalho, Persson, Finstad, and L{\o}vvik}}]{Berland2016:filtering}
\bibinfo{author}{\bibfnamefont{K.}~\bibnamefont{Berland}},
  \bibinfo{author}{\bibfnamefont{X.}~\bibnamefont{Song}},
  \bibinfo{author}{\bibfnamefont{P.~A.} \bibnamefont{Carvalho}},
  \bibinfo{author}{\bibfnamefont{C.}~\bibnamefont{Persson}},
  \bibinfo{author}{\bibfnamefont{T.~G.} \bibnamefont{Finstad}},
  \bibnamefont{and} \bibinfo{author}{\bibfnamefont{O.~M.}
  \bibnamefont{L{\o}vvik}}, \bibinfo{journal}{J. App. Phys.}
  \textbf{\bibinfo{volume}{119}} (\bibinfo{year}{2016}).

\bibitem[{\citenamefont{Chen et~al.}(2013)\citenamefont{Chen, Parker, Du, and
  Singh}}]{example_of_scissor}
\bibinfo{author}{\bibfnamefont{X.}~\bibnamefont{Chen}},
  \bibinfo{author}{\bibfnamefont{D.}~\bibnamefont{Parker}},
  \bibinfo{author}{\bibfnamefont{M.-H.} \bibnamefont{Du}}, \bibnamefont{and}
  \bibinfo{author}{\bibfnamefont{D.~J.} \bibnamefont{Singh}},
  \bibinfo{journal}{New J. Phys.} \textbf{\bibinfo{volume}{15}},
  \bibinfo{pages}{043029} (\bibinfo{year}{2013}).

\bibitem[{\citenamefont{Persson et~al.}(2001)\citenamefont{Persson, Ahuja, and
  Johansson}}]{Persson:2001}
\bibinfo{author}{\bibfnamefont{C.}~\bibnamefont{Persson}},
  \bibinfo{author}{\bibfnamefont{R.}~\bibnamefont{Ahuja}}, \bibnamefont{and}
  \bibinfo{author}{\bibfnamefont{B.}~\bibnamefont{Johansson}},
  \bibinfo{journal}{Phys. Rev. B} \textbf{\bibinfo{volume}{64}},
  \bibinfo{pages}{033201} (\bibinfo{year}{2001}).

\bibitem[{\citenamefont{Persson and Mirbt}(2006)}]{PERSSON2006}
\bibinfo{author}{\bibfnamefont{C.}~\bibnamefont{Persson}} \bibnamefont{and}
  \bibinfo{author}{\bibfnamefont{S.}~\bibnamefont{Mirbt}},
  \bibinfo{journal}{{Braz. J. Phys.}} \textbf{\bibinfo{volume}{36}},
  \bibinfo{pages}{286 } (\bibinfo{year}{2006}).

\bibitem[{\citenamefont{Levine and Allan}(1989)}]{LinOpt}
\bibinfo{author}{\bibfnamefont{Z.~H.} \bibnamefont{Levine}} \bibnamefont{and}
  \bibinfo{author}{\bibfnamefont{D.~C.} \bibnamefont{Allan}},
  \bibinfo{journal}{Phys. Rev. Lett.} \textbf{\bibinfo{volume}{63}},
  \bibinfo{pages}{1719} (\bibinfo{year}{1989}).

\bibitem[{\citenamefont{Del~Sole and Girlanda}(1993)}]{Del_sole:1993}
\bibinfo{author}{\bibfnamefont{R.}~\bibnamefont{Del~Sole}} \bibnamefont{and}
  \bibinfo{author}{\bibfnamefont{R.}~\bibnamefont{Girlanda}},
  \bibinfo{journal}{Phys. Rev. B} \textbf{\bibinfo{volume}{48}},
  \bibinfo{pages}{11789} (\bibinfo{year}{1993}).

\bibitem[{\citenamefont{Armiento and K\"ummel}(2013)}]{AK13}
\bibinfo{author}{\bibfnamefont{R.}~\bibnamefont{Armiento}} \bibnamefont{and}
  \bibinfo{author}{\bibfnamefont{S.}~\bibnamefont{K\"ummel}},
  \bibinfo{journal}{Phys. Rev. Lett.} \textbf{\bibinfo{volume}{111}},
  \bibinfo{pages}{036402} (\bibinfo{year}{2013}).

\bibitem[{\citenamefont{Engel and Vosko}(1993{\natexlab{a}})}]{EngelVosko93}
\bibinfo{author}{\bibfnamefont{E.}~\bibnamefont{Engel}} \bibnamefont{and}
  \bibinfo{author}{\bibfnamefont{S.~H.} \bibnamefont{Vosko}},
  \bibinfo{journal}{Phys. Rev. B} \textbf{\bibinfo{volume}{47}},
  \bibinfo{pages}{13164} (\bibinfo{year}{1993}{\natexlab{a}}).

\bibitem[{\citenamefont{Tran and Blaha}(2009)}]{TranBlaha2009}
\bibinfo{author}{\bibfnamefont{F.}~\bibnamefont{Tran}} \bibnamefont{and}
  \bibinfo{author}{\bibfnamefont{P.}~\bibnamefont{Blaha}},
  \bibinfo{journal}{Phys. Rev. Lett.} \textbf{\bibinfo{volume}{102}},
  \bibinfo{pages}{226401} (\bibinfo{year}{2009}).

\bibitem[{\citenamefont{Gritsenko et~al.}(1995)\citenamefont{Gritsenko, van
  Leeuwen, van Lenthe, and Baerends}}]{GLLB1}
\bibinfo{author}{\bibfnamefont{O.}~\bibnamefont{Gritsenko}},
  \bibinfo{author}{\bibfnamefont{R.}~\bibnamefont{van Leeuwen}},
  \bibinfo{author}{\bibfnamefont{E.}~\bibnamefont{van Lenthe}},
  \bibnamefont{and} \bibinfo{author}{\bibfnamefont{E.~J.}
  \bibnamefont{Baerends}}, \bibinfo{journal}{Phys. Rev. A}
  \textbf{\bibinfo{volume}{51}}, \bibinfo{pages}{1944} (\bibinfo{year}{1995}).

\bibitem[{\citenamefont{Kuisma et~al.}(2010)\citenamefont{Kuisma, Ojanen,
  Enkovaara, and Rantala}}]{GLLB2}
\bibinfo{author}{\bibfnamefont{M.}~\bibnamefont{Kuisma}},
  \bibinfo{author}{\bibfnamefont{J.}~\bibnamefont{Ojanen}},
  \bibinfo{author}{\bibfnamefont{J.}~\bibnamefont{Enkovaara}},
  \bibnamefont{and} \bibinfo{author}{\bibfnamefont{T.~T.}
  \bibnamefont{Rantala}}, \bibinfo{journal}{Phys. Rev. B}
  \textbf{\bibinfo{volume}{82}}, \bibinfo{pages}{115106}
  (\bibinfo{year}{2010}).

\bibitem[{\citenamefont{Shirley}(1996)}]{Shirley}
\bibinfo{author}{\bibfnamefont{E.~L.} \bibnamefont{Shirley}},
  \bibinfo{journal}{Phys. Rev. B} \textbf{\bibinfo{volume}{54}},
  \bibinfo{pages}{16464} (\bibinfo{year}{1996}).

\bibitem[{\citenamefont{Prendergast and Louie}(2009)}]{Predegast:Shirley}
\bibinfo{author}{\bibfnamefont{D.}~\bibnamefont{Prendergast}} \bibnamefont{and}
  \bibinfo{author}{\bibfnamefont{S.~G.} \bibnamefont{Louie}},
  \bibinfo{journal}{Phys. Rev. B} \textbf{\bibinfo{volume}{80}},
  \bibinfo{pages}{235126} (\bibinfo{year}{2009}).

\bibitem[{\citenamefont{Souza et~al.}(2001)\citenamefont{Souza, Marzari, and
  Vanderbilt}}]{wannier2001}
\bibinfo{author}{\bibfnamefont{I.}~\bibnamefont{Souza}},
  \bibinfo{author}{\bibfnamefont{N.}~\bibnamefont{Marzari}}, \bibnamefont{and}
  \bibinfo{author}{\bibfnamefont{D.}~\bibnamefont{Vanderbilt}},
  \bibinfo{journal}{Phys. Rev. B} \textbf{\bibinfo{volume}{65}},
  \bibinfo{pages}{035109} (\bibinfo{year}{2001}).

\bibitem[{\citenamefont{Mostofi et~al.}(2014)\citenamefont{Mostofi, Yates,
  Pizzi, Lee, Souza, Vanderbilt, and Marzari}}]{wannier90new}
\bibinfo{author}{\bibfnamefont{A.~A.} \bibnamefont{Mostofi}},
  \bibinfo{author}{\bibfnamefont{J.~R.} \bibnamefont{Yates}},
  \bibinfo{author}{\bibfnamefont{G.}~\bibnamefont{Pizzi}},
  \bibinfo{author}{\bibfnamefont{Y.-S.} \bibnamefont{Lee}},
  \bibinfo{author}{\bibfnamefont{I.}~\bibnamefont{Souza}},
  \bibinfo{author}{\bibfnamefont{D.}~\bibnamefont{Vanderbilt}},
  \bibnamefont{and} \bibinfo{author}{\bibfnamefont{N.}~\bibnamefont{Marzari}},
  \bibinfo{journal}{Comput. Phys. Commun.} \textbf{\bibinfo{volume}{185}},
  \bibinfo{pages}{2309 } (\bibinfo{year}{2014}).

\bibitem[{\citenamefont{Marzari et~al.}(2012)\citenamefont{Marzari, Mostofi,
  Yates, Souza, and Vanderbilt}}]{WannierReview}
\bibinfo{author}{\bibfnamefont{N.}~\bibnamefont{Marzari}},
  \bibinfo{author}{\bibfnamefont{A.~A.} \bibnamefont{Mostofi}},
  \bibinfo{author}{\bibfnamefont{J.~R.} \bibnamefont{Yates}},
  \bibinfo{author}{\bibfnamefont{I.}~\bibnamefont{Souza}}, \bibnamefont{and}
  \bibinfo{author}{\bibfnamefont{D.}~\bibnamefont{Vanderbilt}},
  \bibinfo{journal}{Rev. Mod. Phys.} \textbf{\bibinfo{volume}{84}},
  \bibinfo{pages}{1419} (\bibinfo{year}{2012}).

\bibitem[{\citenamefont{Pizzi et~al.}(2014)\citenamefont{Pizzi, Volja,
  Kozinsky, Fornari, and Marzari}}]{BoltzWann}
\bibinfo{author}{\bibfnamefont{G.}~\bibnamefont{Pizzi}},
  \bibinfo{author}{\bibfnamefont{D.}~\bibnamefont{Volja}},
  \bibinfo{author}{\bibfnamefont{B.}~\bibnamefont{Kozinsky}},
  \bibinfo{author}{\bibfnamefont{M.}~\bibnamefont{Fornari}}, \bibnamefont{and}
  \bibinfo{author}{\bibfnamefont{N.}~\bibnamefont{Marzari}},
  \bibinfo{journal}{Comput. Phys. Commun.} \textbf{\bibinfo{volume}{185}},
  \bibinfo{pages}{422 } (\bibinfo{year}{2014}).

\bibitem[{\citenamefont{Berland and Persson}(2017)}]{berland201717}
\bibinfo{author}{\bibfnamefont{K.}~\bibnamefont{Berland}} \bibnamefont{and}
  \bibinfo{author}{\bibfnamefont{C.}~\bibnamefont{Persson}},
  \bibinfo{journal}{Comp. Mater. Sci.} \textbf{\bibinfo{volume}{134}},
  \bibinfo{pages}{17 } (\bibinfo{year}{2017}).

\bibitem[{\citenamefont{Persson and Ambrosch-Draxl}(2007)}]{Persson2007280}
\bibinfo{author}{\bibfnamefont{C.}~\bibnamefont{Persson}} \bibnamefont{and}
  \bibinfo{author}{\bibfnamefont{C.}~\bibnamefont{Ambrosch-Draxl}},
  \bibinfo{journal}{Comput. Phys. Commun.} \textbf{\bibinfo{volume}{177}},
  \bibinfo{pages}{280 } (\bibinfo{year}{2007}).

\bibitem[{\citenamefont{Kresse and Hafner}(1993)}]{vasp1}
\bibinfo{author}{\bibfnamefont{G.}~\bibnamefont{Kresse}} \bibnamefont{and}
  \bibinfo{author}{\bibfnamefont{J.}~\bibnamefont{Hafner}},
  \bibinfo{journal}{Phys. Rev. B} \textbf{\bibinfo{volume}{47}},
  \bibinfo{pages}{558} (\bibinfo{year}{1993}).

\bibitem[{\citenamefont{Kresse and Furthmüller}(1996)}]{vasp3}
\bibinfo{author}{\bibfnamefont{G.}~\bibnamefont{Kresse}} \bibnamefont{and}
  \bibinfo{author}{\bibfnamefont{J.}~\bibnamefont{Furthmüller}},
  \bibinfo{journal}{Comput. Mat. Sci.} \textbf{\bibinfo{volume}{6}},
  \bibinfo{pages}{15 } (\bibinfo{year}{1996}).

\bibitem[{\citenamefont{Kresse and Furthm\"uller}(1996)}]{vasp4}
\bibinfo{author}{\bibfnamefont{G.}~\bibnamefont{Kresse}} \bibnamefont{and}
  \bibinfo{author}{\bibfnamefont{J.}~\bibnamefont{Furthm\"uller}},
  \bibinfo{journal}{Phys. Rev. B} \textbf{\bibinfo{volume}{54}},
  \bibinfo{pages}{11169} (\bibinfo{year}{1996}).

\bibitem[{\citenamefont{Gajdo\ifmmode~\check{s}\else \v{s}\fi{}
  et~al.}(2006)\citenamefont{Gajdo\ifmmode~\check{s}\else \v{s}\fi{}, Hummer,
  Kresse, Furthm\"uller, and Bechstedt}}]{vasp:optics}
\bibinfo{author}{\bibfnamefont{M.}~\bibnamefont{Gajdo\ifmmode~\check{s}\else
  \v{s}\fi{}}}, \bibinfo{author}{\bibfnamefont{K.}~\bibnamefont{Hummer}},
  \bibinfo{author}{\bibfnamefont{G.}~\bibnamefont{Kresse}},
  \bibinfo{author}{\bibfnamefont{J.}~\bibnamefont{Furthm\"uller}},
  \bibnamefont{and}
  \bibinfo{author}{\bibfnamefont{F.}~\bibnamefont{Bechstedt}},
  \bibinfo{journal}{Phys. Rev. B} \textbf{\bibinfo{volume}{73}},
  \bibinfo{pages}{045112} (\bibinfo{year}{2006}).

\bibitem[{\citenamefont{Perdew et~al.}(2008)\citenamefont{Perdew, Ruzsinszky,
  Csonka, Vydrov, Scuseria, Constantin, Zhou, and Burke}}]{PBEsol}
\bibinfo{author}{\bibfnamefont{J.~P.} \bibnamefont{Perdew}},
  \bibinfo{author}{\bibfnamefont{A.}~\bibnamefont{Ruzsinszky}},
  \bibinfo{author}{\bibfnamefont{G.~I.} \bibnamefont{Csonka}},
  \bibinfo{author}{\bibfnamefont{O.~A.} \bibnamefont{Vydrov}},
  \bibinfo{author}{\bibfnamefont{G.~E.} \bibnamefont{Scuseria}},
  \bibinfo{author}{\bibfnamefont{L.~A.} \bibnamefont{Constantin}},
  \bibinfo{author}{\bibfnamefont{X.}~\bibnamefont{Zhou}}, \bibnamefont{and}
  \bibinfo{author}{\bibfnamefont{K.}~\bibnamefont{Burke}},
  \bibinfo{journal}{Phys. Rev. Lett.} \textbf{\bibinfo{volume}{100}},
  \bibinfo{pages}{136406} (\bibinfo{year}{2008}).

\bibitem[{\citenamefont{Kane}(1956)}]{KANE1956}
\bibinfo{author}{\bibfnamefont{E.}~\bibnamefont{Kane}}, \bibinfo{journal}{J.
  Phys. Chem. Solids} \textbf{\bibinfo{volume}{1}}, \bibinfo{pages}{82 }
  (\bibinfo{year}{1956}).

\bibitem[{\citenamefont{Luttinger and Kohn}(1955)}]{LuttingerKohn1955}
\bibinfo{author}{\bibfnamefont{J.~M.} \bibnamefont{Luttinger}}
  \bibnamefont{and} \bibinfo{author}{\bibfnamefont{W.}~\bibnamefont{Kohn}},
  \bibinfo{journal}{Phys. Rev.} \textbf{\bibinfo{volume}{97}},
  \bibinfo{pages}{869} (\bibinfo{year}{1955}).

\bibitem[{\citenamefont{Dresselhaus et~al.}(2007)\citenamefont{Dresselhaus,
  Dresselhaus, and Jorio}}]{dresselhaus2007group}
\bibinfo{author}{\bibfnamefont{M.}~\bibnamefont{Dresselhaus}},
  \bibinfo{author}{\bibfnamefont{G.}~\bibnamefont{Dresselhaus}},
  \bibnamefont{and} \bibinfo{author}{\bibfnamefont{A.}~\bibnamefont{Jorio}},
  \emph{\bibinfo{title}{Group Theory: Application to the Physics of Condensed
  Matter}}, SpringerLink: Springer e-Books (\bibinfo{publisher}{Springer},
  \bibinfo{address}{Berlin, Heidelberg}, \bibinfo{year}{2007}).

\bibitem[{\citenamefont{Kim et~al.}(1998)\citenamefont{Kim, Wang, and
  Zunger}}]{Jeongnim1998}
\bibinfo{author}{\bibfnamefont{J.}~\bibnamefont{Kim}},
  \bibinfo{author}{\bibfnamefont{L.-W.} \bibnamefont{Wang}}, \bibnamefont{and}
  \bibinfo{author}{\bibfnamefont{A.}~\bibnamefont{Zunger}},
  \bibinfo{journal}{Phys. Rev. B} \textbf{\bibinfo{volume}{57}},
  \bibinfo{pages}{R9408} (\bibinfo{year}{1998}).

\bibitem[{\citenamefont{von Allmen}(1992)}]{Allmen1992}
\bibinfo{author}{\bibfnamefont{P.}~\bibnamefont{von Allmen}},
  \bibinfo{journal}{Phys. Rev. B} \textbf{\bibinfo{volume}{46}},
  \bibinfo{pages}{15382} (\bibinfo{year}{1992}).

\bibitem[{\citenamefont{Tomi{\'c} and Vukmirovi{\'c}}(2011)}]{Stanko:kp}
\bibinfo{author}{\bibfnamefont{S.}~\bibnamefont{Tomi{\'c}}} \bibnamefont{and}
  \bibinfo{author}{\bibfnamefont{N.}~\bibnamefont{Vukmirovi{\'c}}},
  \bibinfo{journal}{J. App. Phys.} \textbf{\bibinfo{volume}{110}}
  (\bibinfo{year}{2011}).

\bibitem[{\citenamefont{Dresselhaus et~al.}(1955)\citenamefont{Dresselhaus,
  Kip, and Kittel}}]{Dresselhaus1955:kp}
\bibinfo{author}{\bibfnamefont{G.}~\bibnamefont{Dresselhaus}},
  \bibinfo{author}{\bibfnamefont{A.~F.} \bibnamefont{Kip}}, \bibnamefont{and}
  \bibinfo{author}{\bibfnamefont{C.}~\bibnamefont{Kittel}},
  \bibinfo{journal}{Phys. Rev.} \textbf{\bibinfo{volume}{98}},
  \bibinfo{pages}{368} (\bibinfo{year}{1955}).

\bibitem[{\citenamefont{Cardona and Pollak}(1966)}]{Cardona1966}
\bibinfo{author}{\bibfnamefont{M.}~\bibnamefont{Cardona}} \bibnamefont{and}
  \bibinfo{author}{\bibfnamefont{F.~H.} \bibnamefont{Pollak}},
  \bibinfo{journal}{Phys. Rev.} \textbf{\bibinfo{volume}{142}},
  \bibinfo{pages}{530} (\bibinfo{year}{1966}).

\bibitem[{\citenamefont{Voon and Willatzen}(2009)}]{voon2009k}
\bibinfo{author}{\bibfnamefont{L.}~\bibnamefont{Voon}} \bibnamefont{and}
  \bibinfo{author}{\bibfnamefont{M.}~\bibnamefont{Willatzen}},
  \emph{\bibinfo{title}{The {k$\cdot$p} Method: Electronic Properties of
  Semiconductors}} (\bibinfo{publisher}{Springer}, \bibinfo{address}{Berlin,
  Heidelberg}, \bibinfo{year}{2009}).

\bibitem[{\citenamefont{Pickard and Payne}(2000)}]{Pickard2000}
\bibinfo{author}{\bibfnamefont{C.~J.} \bibnamefont{Pickard}} \bibnamefont{and}
  \bibinfo{author}{\bibfnamefont{M.~C.} \bibnamefont{Payne}},
  \bibinfo{journal}{Phys. Rev. B} \textbf{\bibinfo{volume}{62}},
  \bibinfo{pages}{4383} (\bibinfo{year}{2000}).

\bibitem[{\citenamefont{O’Dwyer et~al.}(2006)\citenamefont{O’Dwyer,
  Humphrey, and Linke}}]{InP:nanowire}
\bibinfo{author}{\bibfnamefont{M.~F.} \bibnamefont{O’Dwyer}},
  \bibinfo{author}{\bibfnamefont{T.~E.} \bibnamefont{Humphrey}},
  \bibnamefont{and} \bibinfo{author}{\bibfnamefont{H.}~\bibnamefont{Linke}},
  \bibinfo{journal}{Nanotechnology} \textbf{\bibinfo{volume}{17}},
  \bibinfo{pages}{S338} (\bibinfo{year}{2006}).

\bibitem[{\citenamefont{Zou et~al.}(2015)\citenamefont{Zou, Chen, Huang, Xu,
  and Duan}}]{GaAs:nanowire}
\bibinfo{author}{\bibfnamefont{X.}~\bibnamefont{Zou}},
  \bibinfo{author}{\bibfnamefont{X.}~\bibnamefont{Chen}},
  \bibinfo{author}{\bibfnamefont{H.}~\bibnamefont{Huang}},
  \bibinfo{author}{\bibfnamefont{Y.}~\bibnamefont{Xu}}, \bibnamefont{and}
  \bibinfo{author}{\bibfnamefont{W.}~\bibnamefont{Duan}},
  \bibinfo{journal}{Nanoscale} \textbf{\bibinfo{volume}{7}},
  \bibinfo{pages}{8776} (\bibinfo{year}{2015}).

\bibitem[{\citenamefont{Pichanusakorn et~al.}(2011)\citenamefont{Pichanusakorn,
  Kuang, Patel, Tu, and Bandaru}}]{GaAs:Seebeck}
\bibinfo{author}{\bibfnamefont{P.}~\bibnamefont{Pichanusakorn}},
  \bibinfo{author}{\bibfnamefont{Y.~J.} \bibnamefont{Kuang}},
  \bibinfo{author}{\bibfnamefont{C.~J.} \bibnamefont{Patel}},
  \bibinfo{author}{\bibfnamefont{C.~W.} \bibnamefont{Tu}}, \bibnamefont{and}
  \bibinfo{author}{\bibfnamefont{P.~R.} \bibnamefont{Bandaru}},
  \bibinfo{journal}{App. Phys. Lett.} \textbf{\bibinfo{volume}{99}},
  \bibinfo{pages}{072114} (\bibinfo{year}{2011}).

\bibitem[{\citenamefont{Kesamanly et~al.}(1969)\citenamefont{Kesamanly,
  Nasledov, Nashelskii, and Skripkin}}]{InP:Kesamanly}
\bibinfo{author}{\bibfnamefont{F.~P.} \bibnamefont{Kesamanly}},
  \bibinfo{author}{\bibfnamefont{D.~N.} \bibnamefont{Nasledov}},
  \bibinfo{author}{\bibfnamefont{A.~Y.} \bibnamefont{Nashelskii}},
  \bibnamefont{and} \bibinfo{author}{\bibfnamefont{V.~A.}
  \bibnamefont{Skripkin}}, \bibinfo{journal}{Sov. Phys. Semicond.}
  \textbf{\bibinfo{volume}{2}}, \bibinfo{pages}{1221} (\bibinfo{year}{1969}).

\bibitem[{\citenamefont{Vurgaftman et~al.}(2001)\citenamefont{Vurgaftman,
  Meyer, and Ram-Mohan}}]{Vurgaftman2001}
\bibinfo{author}{\bibfnamefont{I.}~\bibnamefont{Vurgaftman}},
  \bibinfo{author}{\bibfnamefont{J.~R.} \bibnamefont{Meyer}}, \bibnamefont{and}
  \bibinfo{author}{\bibfnamefont{L.~R.} \bibnamefont{Ram-Mohan}},
  \bibinfo{journal}{J. App. Phys.} \textbf{\bibinfo{volume}{89}},
  \bibinfo{pages}{5815} (\bibinfo{year}{2001}).

\bibitem[{\citenamefont{Nolas et~al.}(2013)\citenamefont{Nolas, Sharp, and
  Goldsmid}}]{nolas2013thermoelectrics}
\bibinfo{author}{\bibfnamefont{G.}~\bibnamefont{Nolas}},
  \bibinfo{author}{\bibfnamefont{J.}~\bibnamefont{Sharp}}, \bibnamefont{and}
  \bibinfo{author}{\bibfnamefont{J.}~\bibnamefont{Goldsmid}},
  \emph{\bibinfo{title}{Thermoelectrics: Basic Principles and New Materials
  Developments}}, Springer Series in Materials Science
  (\bibinfo{publisher}{Springer}, \bibinfo{address}{Berlin, Heidelberg},
  \bibinfo{year}{2013}).

\bibitem[{\citenamefont{Sotoodeh et~al.}(2000)\citenamefont{Sotoodeh, Khalid,
  and Rezazadeh}}]{Sotoodeh:2000}
\bibinfo{author}{\bibfnamefont{M.}~\bibnamefont{Sotoodeh}},
  \bibinfo{author}{\bibfnamefont{A.~H.} \bibnamefont{Khalid}},
  \bibnamefont{and} \bibinfo{author}{\bibfnamefont{A.~A.}
  \bibnamefont{Rezazadeh}}, \bibinfo{journal}{J. App. Phys.}
  \textbf{\bibinfo{volume}{87}}, \bibinfo{pages}{2890} (\bibinfo{year}{2000}).

\bibitem[{\citenamefont{Dughaish}(2002)}]{PbTe:review}
\bibinfo{author}{\bibfnamefont{Z.}~\bibnamefont{Dughaish}},
  \bibinfo{journal}{Physica B: Condensed Matter}
  \textbf{\bibinfo{volume}{322}}, \bibinfo{pages}{205 } (\bibinfo{year}{2002}).

\bibitem[{\citenamefont{Heremans et~al.}(2008)\citenamefont{Heremans, Jovovic,
  Toberer, Saramat, Kurosaki, Charoenphakdee, Yamanaka, and
  Snyder}}]{Heremans554}
\bibinfo{author}{\bibfnamefont{J.~P.} \bibnamefont{Heremans}},
  \bibinfo{author}{\bibfnamefont{V.}~\bibnamefont{Jovovic}},
  \bibinfo{author}{\bibfnamefont{E.~S.} \bibnamefont{Toberer}},
  \bibinfo{author}{\bibfnamefont{A.}~\bibnamefont{Saramat}},
  \bibinfo{author}{\bibfnamefont{K.}~\bibnamefont{Kurosaki}},
  \bibinfo{author}{\bibfnamefont{A.}~\bibnamefont{Charoenphakdee}},
  \bibinfo{author}{\bibfnamefont{S.}~\bibnamefont{Yamanaka}}, \bibnamefont{and}
  \bibinfo{author}{\bibfnamefont{G.~J.} \bibnamefont{Snyder}},
  \bibinfo{journal}{Science} \textbf{\bibinfo{volume}{321}},
  \bibinfo{pages}{554} (\bibinfo{year}{2008}).

\bibitem[{\citenamefont{Lee and Mahanti}(2012)}]{Lee:rigid_band}
\bibinfo{author}{\bibfnamefont{M.-S.} \bibnamefont{Lee}} \bibnamefont{and}
  \bibinfo{author}{\bibfnamefont{S.~D.} \bibnamefont{Mahanti}},
  \bibinfo{journal}{Phys. Rev. B} \textbf{\bibinfo{volume}{85}},
  \bibinfo{pages}{165149} (\bibinfo{year}{2012}).

\bibitem[{\citenamefont{Jaworski et~al.}(2011)\citenamefont{Jaworski,
  Wiendlocha, Jovovic, and Heremans}}]{jaworski:alloy}
\bibinfo{author}{\bibfnamefont{C.~M.} \bibnamefont{Jaworski}},
  \bibinfo{author}{\bibfnamefont{B.}~\bibnamefont{Wiendlocha}},
  \bibinfo{author}{\bibfnamefont{V.}~\bibnamefont{Jovovic}}, \bibnamefont{and}
  \bibinfo{author}{\bibfnamefont{J.~P.} \bibnamefont{Heremans}},
  \bibinfo{journal}{Energy Environ. Sci.} \textbf{\bibinfo{volume}{4}},
  \bibinfo{pages}{4155} (\bibinfo{year}{2011}).

\bibitem[{\citenamefont{Pei et~al.}(2011{\natexlab{a}})\citenamefont{Pei,
  LaLonde, Iwanaga, and Snyder}}]{Pei:heavy_hole}
\bibinfo{author}{\bibfnamefont{Y.}~\bibnamefont{Pei}},
  \bibinfo{author}{\bibfnamefont{A.}~\bibnamefont{LaLonde}},
  \bibinfo{author}{\bibfnamefont{S.}~\bibnamefont{Iwanaga}}, \bibnamefont{and}
  \bibinfo{author}{\bibfnamefont{G.~J.} \bibnamefont{Snyder}},
  \bibinfo{journal}{Energy Environ. Sci.} \textbf{\bibinfo{volume}{4}},
  \bibinfo{pages}{2085} (\bibinfo{year}{2011}{\natexlab{a}}).

\bibitem[{\citenamefont{Pei et~al.}(2011{\natexlab{b}})\citenamefont{Pei, Shi,
  LaLonde, Wang, Chen, and Snyder}}]{pei2011convergence}
\bibinfo{author}{\bibfnamefont{Y.}~\bibnamefont{Pei}},
  \bibinfo{author}{\bibfnamefont{X.}~\bibnamefont{Shi}},
  \bibinfo{author}{\bibfnamefont{A.}~\bibnamefont{LaLonde}},
  \bibinfo{author}{\bibfnamefont{H.}~\bibnamefont{Wang}},
  \bibinfo{author}{\bibfnamefont{L.}~\bibnamefont{Chen}}, \bibnamefont{and}
  \bibinfo{author}{\bibfnamefont{G.~J.} \bibnamefont{Snyder}},
  \bibinfo{journal}{Nature} \textbf{\bibinfo{volume}{473}}, \bibinfo{pages}{66}
  (\bibinfo{year}{2011}{\natexlab{b}}).

\bibitem[{\citenamefont{Engel and Vosko}(1993{\natexlab{b}})}]{EngelVosko}
\bibinfo{author}{\bibfnamefont{E.}~\bibnamefont{Engel}} \bibnamefont{and}
  \bibinfo{author}{\bibfnamefont{S.~H.} \bibnamefont{Vosko}},
  \bibinfo{journal}{Phys. Rev. B} \textbf{\bibinfo{volume}{47}},
  \bibinfo{pages}{13164} (\bibinfo{year}{1993}{\natexlab{b}}).

\bibitem[{\citenamefont{Singh}(2010)}]{Singh:PbTe}
\bibinfo{author}{\bibfnamefont{D.~J.} \bibnamefont{Singh}},
  \bibinfo{journal}{Phys. Rev. B} \textbf{\bibinfo{volume}{81}},
  \bibinfo{pages}{195217} (\bibinfo{year}{2010}).

\bibitem[{\citenamefont{Airapetyants et~al.}(1966)\citenamefont{Airapetyants,
  Vinogradova, Dubrovskaya, Kolomoets, and Rudnik}}]{airapetyants1966structure}
\bibinfo{author}{\bibfnamefont{S.}~\bibnamefont{Airapetyants}},
  \bibinfo{author}{\bibfnamefont{M.}~\bibnamefont{Vinogradova}},
  \bibinfo{author}{\bibfnamefont{I.}~\bibnamefont{Dubrovskaya}},
  \bibinfo{author}{\bibfnamefont{N.}~\bibnamefont{Kolomoets}},
  \bibnamefont{and} \bibinfo{author}{\bibfnamefont{I.}~\bibnamefont{Rudnik}},
  \bibinfo{journal}{Sov. Phys. Solid State} \textbf{\bibinfo{volume}{8}},
  \bibinfo{pages}{1069} (\bibinfo{year}{1966}).

\bibitem[{\citenamefont{Gibbs et~al.}(2013)\citenamefont{Gibbs, Kim, Wang,
  White, Drymiotis, Kaviany, and Snyder}}]{PbTe:temp_gap}
\bibinfo{author}{\bibfnamefont{Z.~M.} \bibnamefont{Gibbs}},
  \bibinfo{author}{\bibfnamefont{H.}~\bibnamefont{Kim}},
  \bibinfo{author}{\bibfnamefont{H.}~\bibnamefont{Wang}},
  \bibinfo{author}{\bibfnamefont{R.~L.} \bibnamefont{White}},
  \bibinfo{author}{\bibfnamefont{F.}~\bibnamefont{Drymiotis}},
  \bibinfo{author}{\bibfnamefont{M.}~\bibnamefont{Kaviany}}, \bibnamefont{and}
  \bibinfo{author}{\bibfnamefont{G.~J.} \bibnamefont{Snyder}},
  \bibinfo{journal}{App. Phys. Lett.} \textbf{\bibinfo{volume}{103}},
  \bibinfo{pages}{262109} (\bibinfo{year}{2013}).

\bibitem[{\citenamefont{Pei et~al.}(2012{\natexlab{b}})\citenamefont{Pei,
  LaLonde, Wang, and Snyder}}]{Pei:PbTE_ndoping}
\bibinfo{author}{\bibfnamefont{Y.}~\bibnamefont{Pei}},
  \bibinfo{author}{\bibfnamefont{A.~D.} \bibnamefont{LaLonde}},
  \bibinfo{author}{\bibfnamefont{H.}~\bibnamefont{Wang}}, \bibnamefont{and}
  \bibinfo{author}{\bibfnamefont{G.~J.} \bibnamefont{Snyder}},
  \bibinfo{journal}{Energy Environ. Sci.} \textbf{\bibinfo{volume}{5}},
  \bibinfo{pages}{7963} (\bibinfo{year}{2012}{\natexlab{b}}).

\bibitem[{\citenamefont{Jain et~al.}(2013)\citenamefont{Jain, Ong, Hautier,
  Chen, Richards, Dacek, Cholia, Gunter, Skinner, Ceder et~al.}}]{MatProject}
\bibinfo{author}{\bibfnamefont{A.}~\bibnamefont{Jain}},
  \bibinfo{author}{\bibfnamefont{S.~P.} \bibnamefont{Ong}},
  \bibinfo{author}{\bibfnamefont{G.}~\bibnamefont{Hautier}},
  \bibinfo{author}{\bibfnamefont{W.}~\bibnamefont{Chen}},
  \bibinfo{author}{\bibfnamefont{W.~D.} \bibnamefont{Richards}},
  \bibinfo{author}{\bibfnamefont{S.}~\bibnamefont{Dacek}},
  \bibinfo{author}{\bibfnamefont{S.}~\bibnamefont{Cholia}},
  \bibinfo{author}{\bibfnamefont{D.}~\bibnamefont{Gunter}},
  \bibinfo{author}{\bibfnamefont{D.}~\bibnamefont{Skinner}},
  \bibinfo{author}{\bibfnamefont{G.}~\bibnamefont{Ceder}},
  \bibnamefont{et~al.}, \bibinfo{journal}{APL Mater.}
  \textbf{\bibinfo{volume}{1}} (\bibinfo{year}{2013}).

\end{thebibliography}
\end{document}